\documentclass[article,reprint]{WileyMSP-template}
\usepackage[colorlinks,
linkcolor=blue,
anchorcolor=blue,
citecolor=blue,
]{hyperref}
\usepackage{graphicx}
\usepackage{epstopdf}
\usepackage{breakurl}
\usepackage{multirow}
\usepackage{enumitem}
\usepackage{color}
\usepackage{tablefootnote}
\usepackage{threeparttable}
\usepackage{tabularx}
\usepackage{booktabs}
\usepackage{textcomp}
\usepackage{lineno}
\usepackage{ragged2e}
\begin{document}

\rhead{\includegraphics[width=2.5cm]}

\title{Superconducting Diode Effects: Mechanisms, Materials and Applications}

\maketitle

\author{Jiajun Ma$^{1,2\dag}$}
\author{Ruiya Zhan$^{1,2,3\dag}$}
\author{Xiao Lin$^{1,2*}$}

\begin{affiliations}

$^1$Key Laboratory for Quantum Materials of Zhejiang Province, Department of Physics, School of Science and Research Center for Industries of the Future, Westlake University, Hangzhou 310030, P. R. China. \\ 

$^2$Institute of Natural Sciences, Westlake Institute for Advanced Study, Hangzhou 310024, P. R. China.
\\
$^3$Department of Physics, Fudan University, Shanghai 200438, China.
\\

Email:linxiao@westlake.edu.cn\\

\end{affiliations}


\justifying

\begin{abstract}
\noindent
Superconducting diode effects (SDEs) generally emerge in superconducting systems where both time-reversal and inversion symmetries are broken, showing nonreciprocal current characteristics: nondissipative in one direction and ohmic in the opposite. Since the discovery of the SDEs by Ando $et~al.$ in the noncentrosymmetric superconductor [Nb/V/Ta]$_\mathrm{n}$ in 2020, notable progress has been achieved on both the theoretical and experimental fronts. It has been proposed that intrinsic SDEs are closely linked to various exotic superconducting states, such as the Fulde-Ferrell-Larkin-Ovchinnikov (FFLO) state, topological superconductivity, and chiral superconductivity. Recently, SDEs have emerged as important  experimental tools for detecting symmetry breaking in exotic superconducting states. This advancement not only enhances our understanding of the fundamental nature of SDEs but also opens new possibilities for their applications in superconducting physics and related fields. This review focuses on the recent experimental progress in the observation of the SDEs and discusses their primary mechanisms from the perspective of material properties and symmetry breaking. Finally, we summarize the observed rectification efficiency of SDE devices and discuss future research directions in this rapidly developing field. 
\end{abstract}

\section{INTRODUCTION}

Nonreciprocal transport properties are fundamental to understanding exotic quantum materials and have become a significant focus in condensed matter physics and materials science \cite{rikken1997observation,tokura2018nonreciprocal,cheong2019sos,yan2024structural}. A straightforward example of non-reciprocal charge transport is the p-n junction~\cite{shockley1949theory}, where inversion symmetry breaking (IRS-breaking) at the interface between a p-type and n-type semiconductor induces  directional charge transport, a phenomenon known as the diode effect. 
As the cornerstone of modern electronics, p-n junctions have widespread applications in rectifiers, LEDs, and solar cells. In a similar vein, conductors that exhibit IRS-breaking can enable nonreciprocal charge transport in the presence of magnetic fields. Examples include the interfaces of noncentrosymmetric oxides\cite{choe2019gate,brehin2023gate,zhang2024light}, chiral conductors \cite{rikken2001electrical,seki2016magnetochiral,krstic2002magneto} and organic conductors\cite{pop2014electrical,rikken2022dielectric}. The nonreciprocal transport in these systems is generally attributed to magnetochiral anisotropy (MCA)\cite{rikken2001electrical,wakatsuki2017nonreciprocal,tokura2018nonreciprocal,cheong2019sos}. According to the Onsager reciprocity relations, this effect emerges specifically when both inversion symmetry (IRS) and time-reversal symmetry (TRS) are broken. Nonreciprocal charge transport also occurs in the superconducting fluctuation region of symmetry breaking superconductors\cite{qin2017superconductivity,wakatsuki2017nonreciprocal,itahashi2020nonreciprocal,ando2020observation}. In such systems,  the relevant energy scale is determined by the superconducting energy gap rather than the Fermi energy, resulting in an enhancement of MCA coefficient by several orders of magnitude compared to that in the normal state.

The concept of superconducting diode effects (SDEs) were initially proposed by Hu $et~al.$ in 2007, who considered Josephson junctions (JJs) composed of p-type and n-type superconductors near the superconducting
-Mott insulator transition~\cite{hu2007proposed}. In this configuration, a polar Mott insulator forms at the interface, enabling unidirectional supercurrent flow (also known as Josephson diode effects, JDEs), analogous to that in semiconducting p-n junctions. Despite this early proposal, SDEs were largely overlooked until their solid experimental realization in 2020~\cite{ando2020observation}, when Ando $et~al.$ observed the effects in heterostructures of [Nb/V/Ta]$_\mathrm{n}$ with out-of-plane IRS-breaking. This discovery sparked widespread interest, leading to the observation of SDEs across various superconducting systems, including chiral superconductors~\cite{le2024superconducting,wan2024unconventional}, superconductors with strong spin-orbit coupling (SOC)~\cite{bauriedl2022supercurrent}, JJs~\cite{pal2022josephson,chen2024edelstein,turini2022josephson,baumgartner2022supercurrent,lotfizadeh2024superconducting,costa2023sign,jeon2022zero}, and superconducting films~\cite{hou2023ubiquitous}. The origin underlying SDEs is rather diverse~\cite{PhysRev.135.A550,larkin1965nonuniform,yuan2022supercurrent,daido2022intrinsic,le2024superconducting,sundaresh2023diamagnetic,hou2023ubiquitous,ustavschikov2022diode,moll2023evolution,lyu2021superconducting}, but it has become increasingly clear that breaking both TRS and IRS are generally essential for the emergence of SDEs, except for rare cases of $B$-even SDEs~\cite{wu2022field,liu2024superconducting,qi2025high}. Consequently, SDEs have been proven to be potent tools for probing exotic superconducting states and were recently employed as effective methods to detect chiral superconducting orders~\cite{le2024superconducting,wan2024unconventional}. 

This review aims to examine several archetype SDEs and is organized as follows. First, we discuss three major theoretical models for understanding SDEs. Next, we introduce various systems that exhibit SDEs at finite or zero magnetic fields. Finally, we compile recent experimental data on diode efficiency and evaluate key factors that modulate the SDE performance. As the study of SDEs is still in its early stages, a unified understanding remains elusive. This review may not capture all relevant studies in the rapidly evolving landscape of SDE research.

\section{Theoretical models}
To provide a coherent understanding of this review, we begin by briefly discussing the theoretical models for interpreting SDEs. These models are diverse and can tentatively be divided into three categories: an exotic superconducting order with finite-momentum pairing~\cite{PhysRev.135.A550,larkin1965nonuniform,yuan2022supercurrent,daido2022intrinsic}, asymmetric motion of vortices~\cite{moll2023evolution,lyu2021superconducting,ustavschikov2022diode,hou2023ubiquitous,sundaresh2023diamagnetic,PhysRevB.72.064509} and nontrivial current-phase relation (CPR) in JJs~\cite{PhysRevLett.129.267702,costa2023sign,li2024interfering,reinhardt2024link,strambini2020josephson}. Although a unified picture underlying SDEs has yet to be established,  there is a general consensus that the emergence of SDEs necessitates the breaking of both TRS and IRS. Note that the observation of $B$-even zero-field SDEs in polar systems challenges current understanding and warrants further investigation~\cite{wu2022field,liu2024superconducting}.

\subsection{Finite-momentum Cooper pairing}
In two-dimensional systems with out-of-plane IRS-breaking, Rashba SOC induces band splitting, resulting in nontrivial spin textures at the Fermi surface. When an in-plane magnetic field ($B_\mathrm{in}$) is applied, additional band shifts occur. Electron pairing at this Fermi surface generates finite-momentum Cooper pairs (seen in Figure~\ref{1}a) with the phase of the order parameter modulated in real space $\Delta({r})=\Delta {e}^{iq\cdot{r}}$, where $q$ is the momentum determined by external $B_\mathrm{in}$~\cite{agterberg2007magnetic, akbari2022fermi}. This state is phenomenologically akin to the Fulde-Ferrell-Larkin-Ovchinnikov (FFLO) state \cite{PhysRev.135.A550,larkin1965nonuniform}. Building on this framework, Yuan $et~al.$ predicted that the pair-breaking effect varies, depending on whether the supercurrent flows parallel or antiparallel to the direction of $q$~\cite{yuan2022supercurrent}. This directional dependence is manifested as SDEs in the transport properties (Figure~\ref{1}b). By utilizing the Ginzburg-Landau (GL) theory, Daido $et~al.$ reveals that the amplitude of asymmetric critical current follows $\Delta {j_c} \propto (T_\mathrm{c}-T)^2$ near the critical temperature ($T_\mathrm{c}$)~\cite{daido2022intrinsic}. And the SDE polarity shows sign reversals under moderate and high fields, which may serve as a signature for intrinsc SDEs. Moreover, Ili\'{c} $et~al.$ predicted that SDEs persist in strongly disordered Rashba superconductors, with the polarity reversing as $B$ and disorder strengthen~\cite{ilic2022theory}. Theories have also discussed higher-order effects in Rashba supercondutors in generating SDEs~\cite{daido2022intrinsic,hasan2024supercurrent,daido2022superconducting,he2022phenomenological}.

\subsection{Asymmetric vortex dynamics}
In 1952, Abrikosov predicted that in Type-II superconductors, magnetic field enters into the superconductor in form of quantized vortices~\cite{abrikosov2017fundamentals}. The moving vortices tend to be pinned by defects in superconductors from an energetic perspective, which helps lower the free energy of the entire system. When a current flows, the vortex would experience a Lorentz force, causing it to move perpendicular to the direction of the current. This perpendicular motion generates an electric field aligned with the current, therefore disrupting the zero-resistance state, and leading to energy dissipation in the form of heat. If the dissipation associated with vortex motion is asymmetric with respect to the direction of current flow, it gives rise to SDEs. In this scenario, the vortex dynamics is determined by the asymmetric potential generated by asymmetrical boundary~\cite{margineda2023sign,moll2023evolution,hou2023ubiquitous,PhysRevApplied.22.064017}, defect organization~\cite{lyu2021superconducting,PhysRevB.110.174509}, and surface conditions~\cite{li2024unconventional,jiang2022field,sundaresh2023diamagnetic}.

\begin{figure*}[t]
    \centering
    \includegraphics[width=1\linewidth]{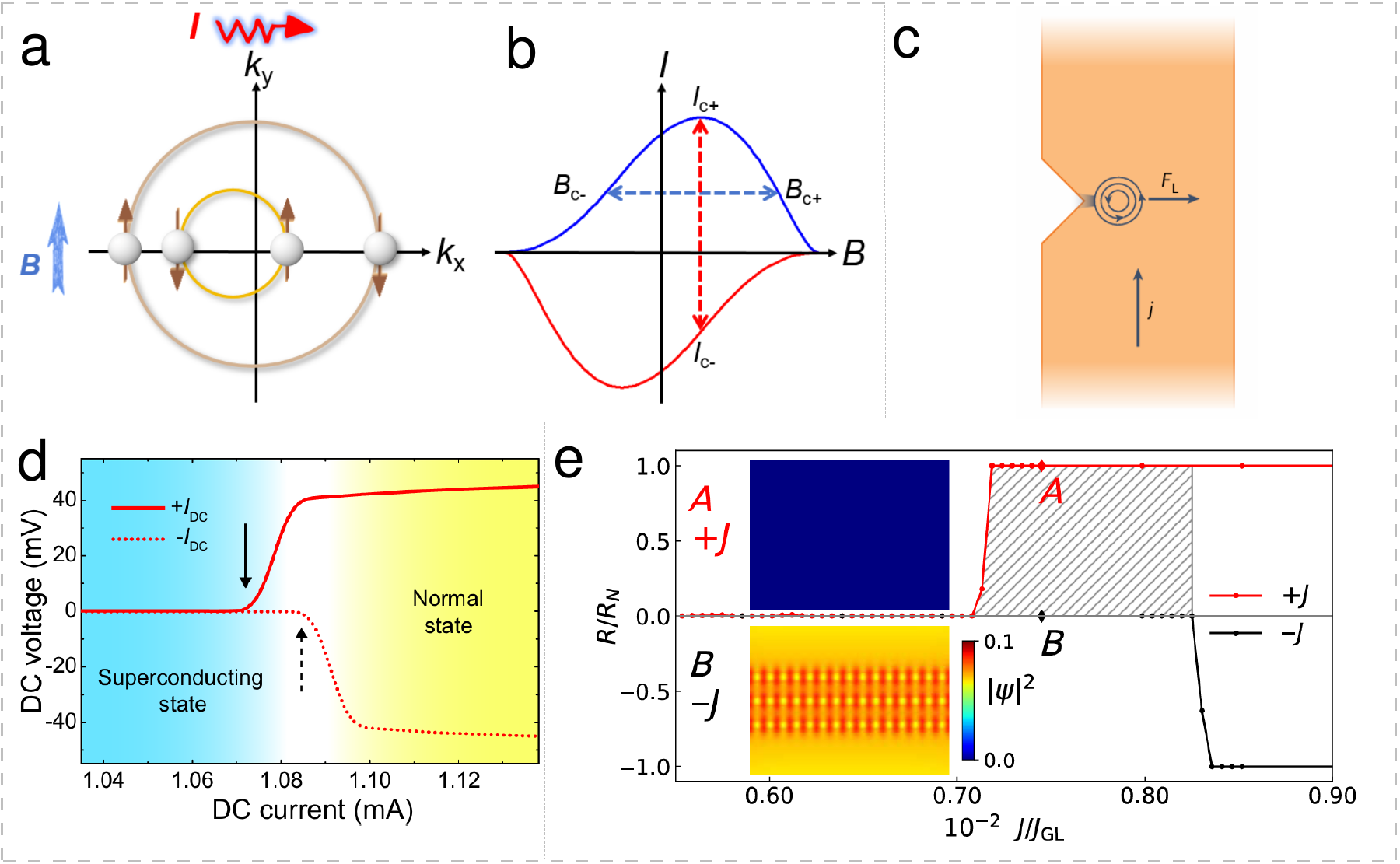}
    \caption{\textbf{Mechanism of SDEs: finite-momentum Cooper pairing and vortex motion.} a) Schematic illustration of finite-momentum Cooper pairing on the Fermi surface featuring nontrivial spin textures, induced by Rashba SOC, under the influence of $B_\mathrm{in}$. b) Illustration of asymmetric critical current and critical field. c) Schematic of vortex entry across a boundary defect. $F_\mathrm{L}$ is the Lorentz force. d) Asymmetric critical current in a MoGe microbridge patterned with conformal nanoholes. e) Simulations of vortex motion-induced SDEs in superconducting films incorporating arrays of magnet dots. (c) is reproduced with permission~\cite{moll2023evolution}, Copyright 2023, Springer Nature Limited. (d) is reproduced with permission~\cite{lyu2021superconducting}, Copyright 2021, The Authors. (e) is reproduced with permission~\cite{jiang2022field}, Copyright 2022, American Physical Society.}
    \label{1}
\end{figure*}

In the context, the initial entry of vortices into the superconductor is significantly affected by asymmetrical vortex edges and surface barriers, which play a crucial role in the nonreciprocal behavior of critical supercurrents, as shown in Figure~\ref{1}c~\cite{moll2023evolution}. Wang $et~al.$ showed that SDEs can be induced by antisymmetric vortex entry, which arises from asymmetric surface thickness on a stepped NbSe$_2$ flake~\cite{PhysRevApplied.22.064017}. Moreover, Lyu $et~al.$ demonstrated significant SDEs in a conventional superconducting film (seen in Figure~\ref{1}d) patterned with a conformal array of nanoscale holes, which creates an asymmetric vortex pinning potential~\cite{lyu2021superconducting}. Furthermore, Du $et~al.$ reported artificially etched $T_\mathrm{d}$-MoTe$_2$ thin films can introduce defects, which can lead to the formation of asymmetric pinning potentials and the generation of SDEs~\cite{PhysRevB.110.174509}. Hou $et~al.$ proposed a simplified screen current model to account for the observed high rectification efficiency of extrinsic SDEs in superconducting stripes~\cite{hou2023ubiquitous,sundaresh2023diamagnetic}. Other studies have explored the possibility in the enhancing SDEs by integrating superconducting films with magnetic stripes~\cite{PhysRevB.72.064509,jiang2022field,ustavschikov2022diode}. Recently, Jiang $et~al.$ theoretically investigated a superconducting film with periodically organized magnetic dots, where the activated vortex-antivortex pairs exhibited asymmetric behavior in their nucleation, movement, and annihilation relative to the current direction, resulting in a pronounced diode response at zero field~\cite{jiang2022field}, as depicted in Figure~\ref{1}e.

The vortex-induced SDEs are typically extrinsic and irrelevant to intrinsic superconducting properties. Therefore, their widespread presence may hinder the search for the intrinsic superconducting orders characterized by finite-momentum Cooper pairing~\cite{hou2023ubiquitous}. While from a functional perspective, this mechanism offers an effective knob to manipulate the diode efficiency through structure engineering and optimization, laying the foundation for future applications~\cite{lyu2021superconducting}.

\subsection{Unconventional current-phase relation}

In JJs, the relationship between the supercurrent and the phase difference across  two superconductors is known as CPR~\cite{golubov2004current}, given by ${I_\mathrm{s}=I_\mathrm{c}\sin(\varphi)}$. For standard JJs, the supercurrent follows a sinusoidal function, vanishing at $\varphi=0$. 
While considering the presence of Rashba SOC and a Zeeman field~\cite{PhysRevLett.101.107005,bergeret2015theory,bezuglyi2002combined,zazunov2009anomalous,yokoyama2014anomalous}, an anomalous Josephson coupling with a non-sinusoidal CPR and a finite phase ($\varphi_0$) shift occurs. In this case, the free energy of the system takes the following form\cite{bergeret2015theory,strambini2020josephson}: $F_L{\sim}f(\alpha,h)(\mathbf{n_{h}}\times\hat{\mathbf{z}})\cdot\mathbf{v_{s}}$, where $f(\alpha,h)$ is a function of the strength of the Rashba SOC ($\alpha$) and the Zeeman field (${h}$), $\mathbf{n_{h}}$ denotes the direction of the Zeeman field, and $\mathbf{v_{s}}$ represents the  supercurrent velocity. Notably, this scalar product reveals asymmetric properties related to the supercurrent direction. In a complementary approach, Davydova $et~al.$ put forth a simple approach to achieve finite-momentum pairing based on the Meissner effect, independent of SOC. They suggested a universal mechanism for SDEs in short JJs, in which the diode effects arise from the Doppler shift of the Andreev bound state energies for finite-momentum Cooper pairs~\cite{davydova2022universal}.

Phenomenologically, incorporating the second-order harmonic term suffices to describe the essential CPR for most cases: $I_\mathrm{s}(\varphi)\approx I_\mathrm{c1}\sin(\varphi+\varphi_0)+I_\mathrm{c2}\sin(2\varphi)$. This relation could yield $I_\mathrm{c+} (\varphi) \ne I_\mathrm{c-}(\varphi)$ at ambient field, leading to SDEs. This phenomenon is also referred to as the JDEs, since it occurs in JJs. Note that JDEs can also arise from extrinsic space/time asymmetry, which result from the combined effects of self-field effects from nonuniform bias and stray fields from a trapped Abrikosov vortex in conventional planar JJs~\cite{golod2022demonstration}.

\section{SDEs under finite magnetic field}

\subsection{SDEs in heterostructures}

\begin{figure*}[t]
    \centering
    \includegraphics[width=1\linewidth]{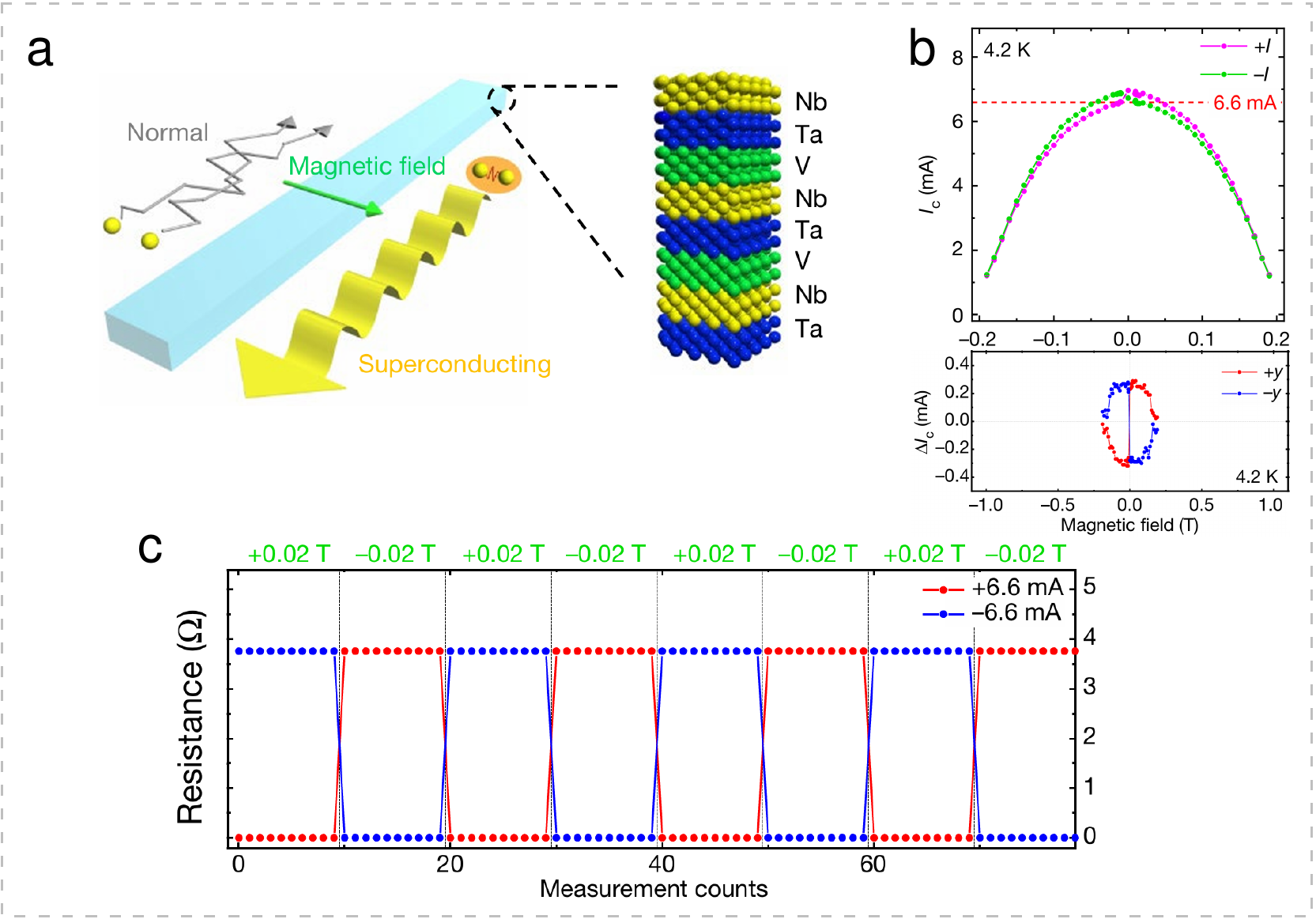}
    \caption{\textbf{SDEs in the [Nb/V/Ta]$_\mathrm{n}$ heterostructure.} a) Schematic illustration of the SDEs manipulated by an in-plane field in the artificial [Nb/V/Ta]$_\mathrm{n}$ superlattice.     
    b) Field dependence of asymmetric $I_\mathrm{c}$ and $\Delta I_\mathrm{c}$. c) Half-wave rectification effects observed by switching the direction of currents or fields. The figures are reproduced with permission~\cite{ando2020observation}, Copyright 2020, The Authors.}
    \label{2}
\end{figure*}

Tracing back to 2009, a subtle signature of a bistable superconducting state related to the direction of current flow was observed in Py/Nb (ferromagnet/superconductor) bilayers~\cite{carapella2009bistable}. This phenomenon was attributed to a unique distribution of screening currents, resulting in vortex diode effects~\cite{gutfreund2023direct}. The solid experimental demonstration of SDEs was reported by Ando $et~al.$ in 2020, who investigated a noncentrosymmetric superlattice composed of alternating layers of three superconducting materials-Nb, V, and Ta, as illustrated in Figure~\ref{2}a~\cite{ando2020observation}. In this heterostructure, IRS is broken along the c-axis. When $B_\mathrm{in}$ is applied, a notable asymmetry in $I_\mathrm{c}$ appears in two opposite current directions (normal to the field direction), and SDE polarity, defined as the sign of $\Delta I_\mathrm{c}$~(=$|I_\mathrm{c+}|-|I_\mathrm{c-}|$), is antisymmetric with respect to $B_\mathrm{in}$, dubbed as $B$-odd SDEs, as seen in Figure~\ref{2}b. 
By setting the driving current between $|I_\mathrm{c+}|$ and $|I_\mathrm{c-}|$, a superconducting rectification effect is observed when the current direction is reversed, as illustrated in Figure~\ref{2}c. The mechanism underlying this phenomenon is likely associated with pairing between Rashba-split bands induced by the out-of-plane IRS-breaking.

\subsection{SDEs in Josephson junctions}

\begin{figure*}[h]
    \centering
    \includegraphics[width=1\linewidth]{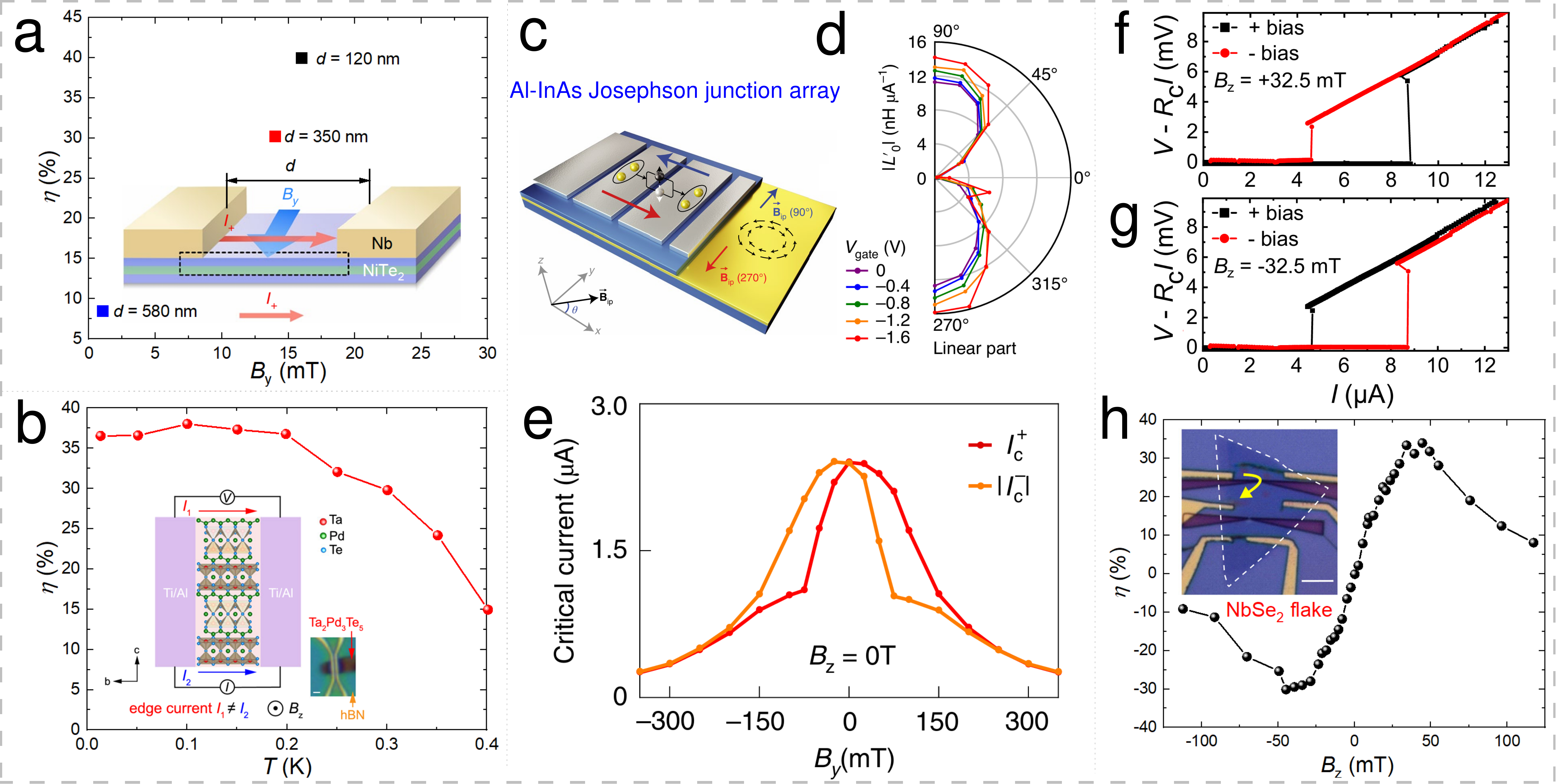}
    \caption{\textbf{SDEs in JJs and homostructures.} a) Diode efficiency in NiTe$_2$-based JJs under finite magnetic fields, shown for different electrode separations. The inset illustrates the device structure with superconducting Nb contacts. b) Temperature dependence of diode efficiency in Ta$_2$Pd$_3$Te$_5$-based JJs, with the device schematic (inset) featuring Ti/Al contacts. c) Schematic illustration of Al-InAs JJs. d) Angle dependence of the Josephson inductance slope ($|L^{'}_{0}|$) in Al-InAs JJs, measured at varying gate voltages. The angle denotes the orientation between $B_\mathrm{in}$ and the current flow, with current applied along the $x$-axis. e) Asymmetric $I_\mathrm{c}$ as a function of $B_\mathrm{y}$ in Al-InAs JJs. f,g) Asymmetric current-voltage characteristics in NbSe$_2$ flakes under out-of-plane $B_\mathrm{z} = \pm32.5~\mathrm{mT}$, respectively. h) $B_\mathrm{z}$-dependent diode efficiency in NbSe$_2$ flakes, highlighting $B$-odd diode behavior.   
    (a) is reproduced with permission~\cite{pal2022josephson}, Copyright 2022, The Authors. (b) is reproduced with permission~\cite{li2024interfering}, Copyright 2024, The Authors. (c-e) are reproduced with permission~\cite{baumgartner2022supercurrent}, Copyright 2022, The Authors. (f-h) are reproduced with permission~\cite{bauriedl2022supercurrent}, Copyright 2022, The Authors, under exclusive license to Springer Nature Limited.}
    \label{3}
\end{figure*}

JDEs have been investigated in JJs that incorporate topological semimetals as barriers. For instance, JJs featuring NiTe$_2$, a type-II Dirac semimetal, as barriers demonstrate JDEs with approximately 40\% diode efficiency, defined as $\eta$ = $(I_{c+}-\left|I_{c-}\right|)/(I_{c+}+\left|I_{c-}\right|)\times100\%$, under $B_\mathrm{in}$ of 12~mT, as illustrated in Figure~\ref{3}a~\cite{pal2022josephson}. The authors attributed this phenomenon to the potential finite-momentum Cooper pairing induced by the applied $B_\mathrm{in}$ within the spin-polarized Fermi surface of NiTe$_2$, as revealed through the combined analysis of angle-resolved photoemission spectroscopy  and density-functional calculations. The similar mechanism was also employed to account for JDEs observed in the Dirac semimetal 1$T$-PtTe$_2$~\cite{ParkinCommunicationsPhysics2024}. Additionally, JDEs with diode efficiency as large as 50.4\% were observed in $T_\mathrm{d}$-MoTe$_2$ based JJs under finite out-of-plane magnetic field ($B_\mathrm{out}$)~\cite{chen2024edelstein}. The distorted MoTe$_2$ is a type-II Weyl semimetal~\cite{keum2015bandgap}, which was predicted to harbor the Edelstein effects, potentially resulting in nontrivial CPR in the junctions as discussed by the authors. However, in the presence of $B_\mathrm{out}$, one should be cautious about extrinsic origins stemming from vortex dynamics and nonuniform bias in inducing JDEs before reaching a definitive conclusion~\cite{golod2022demonstration}. Another type-II Weyl semimetal $T_\mathrm{d}$-WTe$_2$, with an in-plane polarization axis, was employed by Kim $et~al.$ to  fabricate NbSe$_2$/WTe$_2$/NbSe$_2$ vertical JJs for the investigation of JDEs. They observed a sinusoidal dependence of JDEs by the rotating the in-plane magnetic field, highlighting the magnetic chirality effect~\cite{kim2024intrinsic}. Recently, Li $et~al.$ reported the observation of JDEs with diode efficiency reaching 36.5\% in JJs composed of topological insulator Ta$_2$Pd$_3$Te$_5$ and superconducting Al under external  $B_\mathrm{out}$, as illustrated in Figure~\ref{3}b~\cite{li2024interfering}. They attributed the JDEs to anomalous CPR resulting from the combined effects of non-zero magnetic flux, unequal supercurrent induced by the asymmetric edge states and higher harmonics in the edge channels~\cite{PhysRevLett.129.267702}. Overall, Topological JJs with their unique properties such as spin-polarized surface states, strong SOC, and exotic edge states, hold significant potential for advancing research in SDEs.

In addition to topological physics, Rashba SOC in inversion-asymmetric systems can also contribute to the development of JDEs~\cite{baumgartner2022supercurrent,turini2022josephson,lotfizadeh2024superconducting,costa2023sign}, as discussed previously. In 2022, Baumgartner $et~al.$ reported the investigation of nontrivial CPR through Josephson inductance measurements, elucidating its connection to JDEs in a highly transparent Al-InAs Josephson junction array, as illustrated in Figure~\ref{3}c~\cite{baumgartner2022supercurrent}. They explored the angle dependence of Josephson inductance slope by varying the direction of $B_\mathrm{in}$, as illustrated in Figure~\ref{3}d. The slope shows a maximum when $B_\mathrm{in}$ is applied along the $y$ axis (90$^\mathrm{o}$, normal to the current direction).  It leads to the emergence of JDEs in Figure~\ref{3}e, underscoring the significant role of Rashba SOC. The performance of JDEs relies on the mutual configuration of $B_\mathrm{in}$ and current~\cite{lotfizadeh2024superconducting}. Similar findings were reported in InSb nanoflake-based JJs with strong Rashba SOC~\cite{turini2022josephson}.

\subsection{SDEs in homostructures}
SDEs were also observed in constrained few-layer NbSe$_2$, as shown in Figure~\ref{3}f-h~\cite{bauriedl2022supercurrent}. By applying a moderate $B_\mathrm{out}$, the diode rectification efficiency as high as 30\% was demonstrated, as shown in Figure~\ref{3}h. $\Delta I_\mathrm{c}$ exhibits typical antisymmetric behavior as a function of $B_\mathrm{out}$, however, such behavior is disrupted in the presence of additional $B_\mathrm{in}$. The authors attributed the observed SDEs to the valley-Zeeman SOC specific to transition metal dichalcogenides. 
It is important to note that vortex effects are not easy to be excluded when $B_\mathrm{out}$ is applied. Recent work by Hou $et~al.$ reported similar SDEs with large $\eta$ of approximately 65\%  in elemental superconducting  films (V and Nb)~\cite{hou2023ubiquitous}. They suggested that the origin of the SDEs is primarily  extrinsic, arising from asymmetric vortex edge, surface barriers and the associated Meissner screening current. Such extrinsic effects must be carefully considered and eliminated in the search for exotic intrinsic superconducting orders.

\section{SDEs under zero magnetic field}

As discussed above, SDEs are a consequence of simultaneously breaking of IRS and TRS. ISR-breaking could be intrinsic, resulting from lattice effects, or extrinsic, arising from asymmetric geometries. TRS, on the other hand, could be broken by an external $B$ or internal mechanisms, with the later leading to zero-filed SDEs. 

\subsection{Zero-field SDEs in magnetic heterostructures}

\begin{figure*} [h]
    \centering
    \includegraphics[width=1\linewidth]{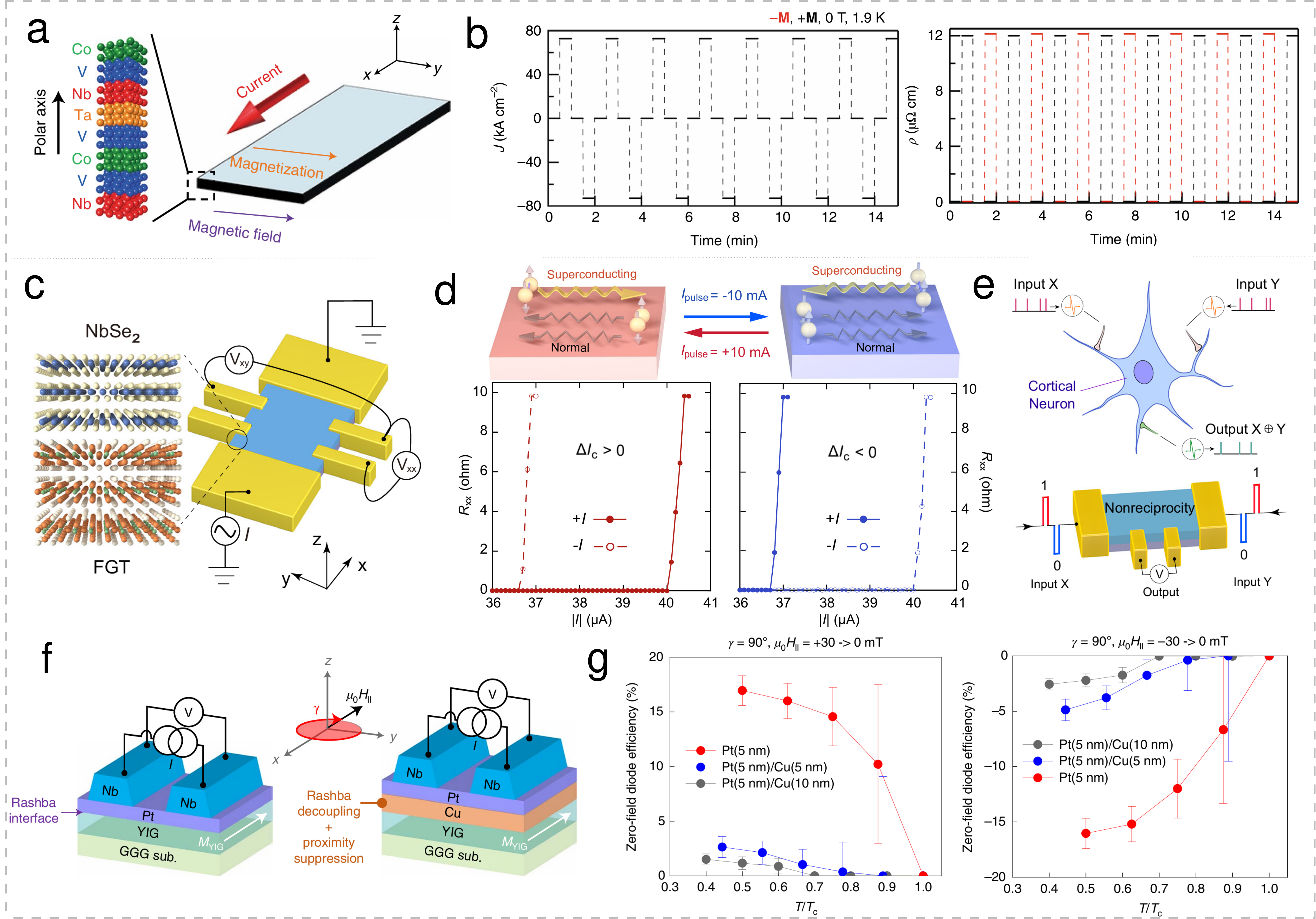}
      \caption{\textbf{Zero-field SDEs in magnetic heterostructures.} a) Schematic of [Nb/V/Co/V/Ta]$_{20}$ multilayers. b) Application of an alternating current (left panel) and the corresponding switchable zero-field SDE characteristics (right panel), demonstrating non-volatile, reversible polarity trained by an external magnetic field. c) Schematic of SDE device based on an NbSe$_2$/Fe$_3$GeTe$_2$ heterostructure. d) Asymmetric $I_\mathrm{c}$ measured in the device from (c), indicating electrically switchable polarity. e) Neuromorphic application concept based on NbSe$_2$/Fe$_3$GeTe$_2$ devices. f) Schematics of Pt/Y$_3$Fe$_5$O$_{12}$ and Pt/Cu/Y$_3$Fe$_5$O$_{12}$ based junctions. g) Zero-field diode efficiency as a function of normalized temperature $T/T_\mathrm{c}$ with varying Cu layer thickness. Its polarity is controlled by magnetization orientation trained by external magnetic field. (a,b) are reproduced with permission~\cite{narita2022field}, Copyright 2022, The Authors, under exclusive licence to Springer Nature Limited. (c-e) are reproduced with permission~\cite{xiong2024electrical}, Copyright 2024, The Authors. (f,g) are reproduced with permission~\cite{jeon2022zero}, Copyright 2022, The Authors, under exclusive licence to Springer Nature Limited.}
    \label{4}
\end{figure*}

Following the discovery of SDEs in [Nb/V/Ta]$_\mathrm{n}$ heterostructure by Ando $et~ al.$ ~\cite{ando2020observation}, the same group introduced an additional layer of cobalt (Co) into the artificial superlattice, constructing [Nb/V/Co/V/Ta]$_{20}$ multilayers, as shown in Figure~\ref{4}a~\cite{narita2022field}. In this configuration, the Co layer exhibits bulk in-plane magnetization, serving as an internal source of TRS-breaking. A pronounced signature of asymmetric $I_\mathrm{c}$ was detected in this heterostructure, enabling rectification without the need for external $B$. By manipulating the magnetization direction of the ferromagnetic layers through an external $B$, zero-field SDE polarity could be effectively reversed, as seen in Figure~\ref{4}b. SDEs could be tuned in a controllable manner by adjusting structural parameters such as constituent elements, film thickness, stacking order and number of repetitions. Building on this idea, similar effects were identified in Fe/Pt-inserted noncentrosymmetric [Nb/V/Ta]$_\mathrm{n}$ superlattices, where the Fe/Pt layer functions as a ferromagnetic component layer~\cite{narita2023magnetization}. The controlled fabrication of field-free SDEs with zero energy loss in heterostructures may enable novel non-volatile memories and logic circuits with ultralow power consumption.

Recently, Xiong $et~al.$ reported the first experimental demonstration of field-free non-volatile electrically switchable SDEs in a van der Waals(vdW) heterostructure composed of NbSe$_2$ and ferromagnetic Fe$_3$GeTe$_2$, as illustrated in Figure~\ref{4}c
~\cite{xiong2024electrical}. Similar work has been reported by ref.~\citenum{hu2025tunable,yun2023magnetic}. NbSe$_2$ is known as an Ising superconductor and the magnetic proximitiy effect from Fe$_3$GeTe$_2$, characterized by out-of-plane magnetic order, leads to valley-contrasting Ising SOC that breaks TRS. This mechanism is similar to valley-Zeeman splitting induced by external $B$ in the single NbSe$_2$ device discussed previously. The spatial asymmetry, essential for SDEs, arises from the breaking of mirror symmetry, caused by the inherent strain and lattice mismatch at the vdW interface or by the interplay of valley-contrasting trigonal warping in NbSe$_2$. The combination of these symmetry-breaking mechanisms is considered to facilitate the emergence of zero-field SDEs. The electrical manipulation of zero-field SDE polarity stems from reversing the magnetization through current induced out-of-plane spin accumulation, as illustrated in Figure.~\ref{4}d. Based on this observation, the authors present a proof-of-concept nonreciprocal quantum neuronal transistor that emulates the biological functionality of cortical neurons in the brain, as depicted in Figure.~\ref{4}e.

\subsection{Zero-field SDEs in JJs}

Zero-field JDEs have been demonstrated in JJs with magnetic Rashba-type barriers. Jeon $et~al.$ reported zero-field polarity-switchable SDEs with maximal efficiency of 35\% at 2~K in Nb-Pt-Nb JJs fabricated on top of ferrimagnetic insulating Y$_3$Fe$_5$O  (YIG) layer, as shown in Figure \ref{4}f~\cite{jeon2022zero}. In this configuration, the Pt/YIG heterojunction serves as the Josephson barrier. Moreover, the zero-field JDE performance could be affected by the presence of an additional Cu buffer layer between Pt and YIG, as illustrated in Figure \ref{4}g, which is believed to affect the interfacial properties. Therefore, the exchange spin-splitting and Rashba SOC at the Pt/YIG interface were considered as key factors in the observation of zero-field JDEs. Note that in superconductor-ferromagnet heterojunctions, the exchange field from the ferromagnet modifies the quasiparticle distribution in the superconductor, potentially leading to exotic superconducting orders such as spin-triplet pairing states~\cite{robinson2010controlled} and the FFLO state~\cite{PhysRev.135.A550,larkin1965nonuniform}. In the context of JJs, these phases could result in $\varphi_0$-junctions, response for zero-field SDE functionality. Zero-field SDEs have also been observed in vdW JJs based on iron-based superconductors Fe(Te,Se)~\cite{qiu2023emergent,li2024field}. Qiu $et~al.$ attributed the phenomenon to the emergence of interfacial ferromagnetism in the superconducting state of Fe(Te,Se)~\cite{qiu2023emergent}. It's worth noting that the zero-field JDEs can also be realized in Nb planar JJs due to a combination of self-field effect and stray field from a trapped Abrikosov vortex~\cite{golod2022demonstration}.

\subsection{Zero-field SDEs from unconventional superconductivity}

\begin{figure*}[t]
    \centering
    \includegraphics[width=1\linewidth]{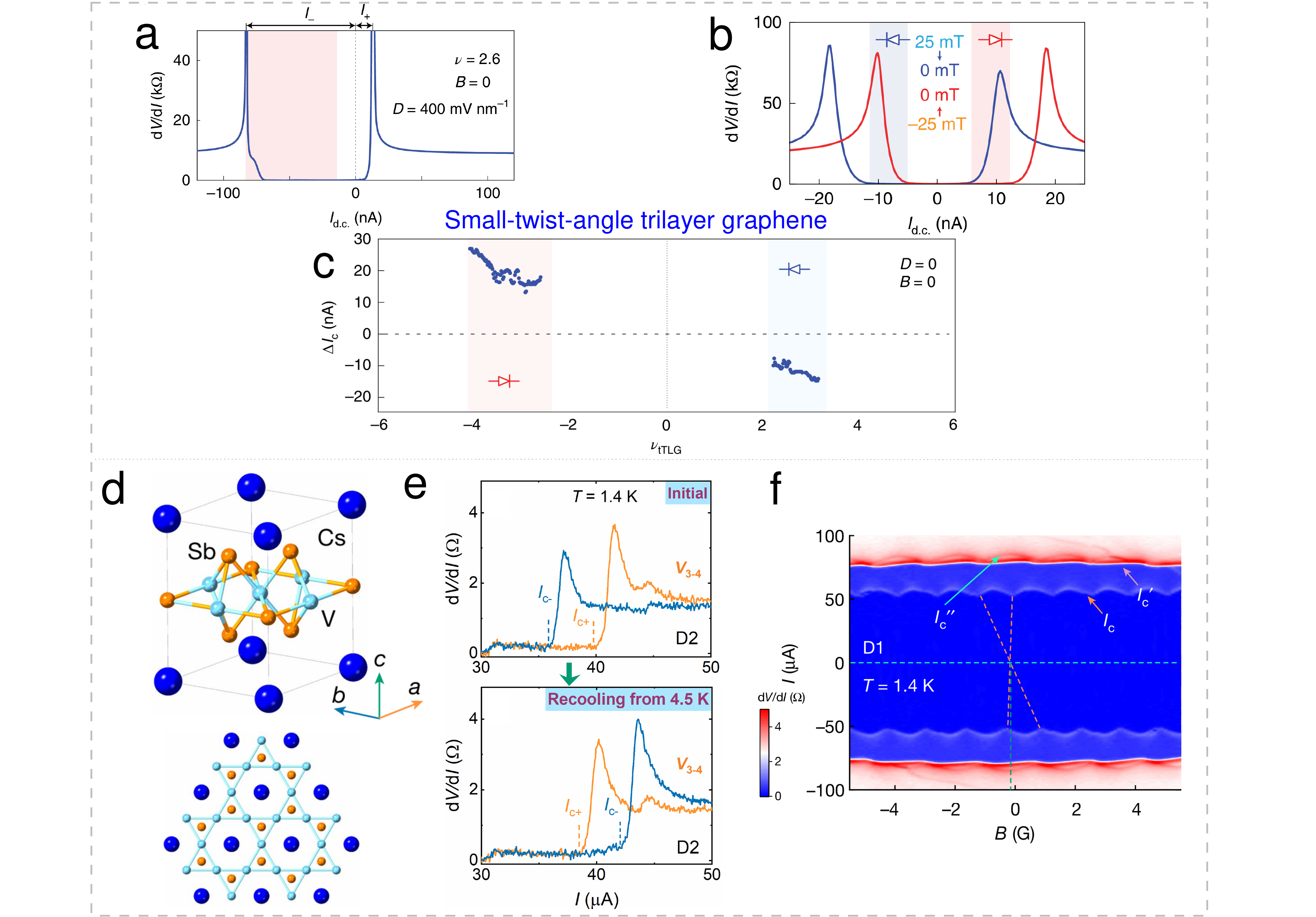}
    \caption{\textbf{Zero-Field SDEs in intrinsic flakes.} a) Differential resistance (d$V$/d$I$) as a function of current bias in tTLG at $B=0,~T=20~\mathrm{mK}$ after positive field training. b) Reservable diode polarity achieved by training with positive and negative field, respectively. c) $\Delta I_\mathrm{c}$ versus carrier filling ($\nu_\mathrm{tTLG}$), showing electrostatic gate modulation.
    d) Crystal structure of CsV$_3$Sb$_5$. e) Zero-field SDEs in intrinsic CsV$_3$Sb$_5$ flakes, with polarity tuned  thermal cycling. f) Quantum interference patterns observed in CsV$_3$Sb$_5$ flakes.
    (a-c) are reproduced with permission~\cite{lin2022zero}, Copyright 2022, The Authors, under exclusive licence to Springer Nature Limited. (d-f) are reproduced with permission~\cite{le2024superconducting}, Copyright 2024, The Authors, under exclusive licence to Springer Nature Limited.}
    \label{5}
\end{figure*}

Zero-field SDEs could serve as powerful tools to detect intrinsic TRS-breaking, reflecting the exotic nature of pairing symmetry in unconventional superconductors. In small-twist angle trilayer graphene (tTLG), remarkable SDEs were observed at zero external magnetic field, as shown in Figure \ref{5}a~\cite{lin2022zero}. Their polarities can be non-volatilely reversed by training with $B_\mathrm{out}$ in Figure \ref{5}b. In addition, the characteristics of the diode effects can be further controlled by varying the carrier concentration through electrostatic gating (seen in Figure \ref{5}c) and twist angle engineering. A potential interpretation for the origin of these effects lies in the unbalanced valley occupancy at the Fermi surface, which may lead to finite-momentum Cooper pairing in the superconducting state. Similar results have also been reported in magic-angle twisted bilayer graphene. \cite{diez2023symmetry}.

Recently, zero-field SDEs have been employed to detect unconventional chiral superconducting orders. Le. el al. reported zero-field SDEs with properties modulated  by thermal cycling in exfoliated CsV$_3$Sb$_5$ flakes, a V-based kagome system ~\cite{le2024superconducting}, as seen in Figure \ref{5}d,e. In conjunction with quantum interference transport measurements in Figure \ref{5}f, this phenomenon is ascribed to the presence of dynamic superconducting domains with TRS-breaking, thereby highlighting the existence of chiral superconducting orders in this system. Additionally, zero-field SDEs were detected in a chiral molecule-intercalated TaS$_2$ hybrid system, providing an important indication of potential chiral superconductivity~\cite{wan2024unconventional}.

\subsection{Zero-field SDEs in polar systems}

\begin{figure*}[h]
    \centering
    \includegraphics[width=1\linewidth]{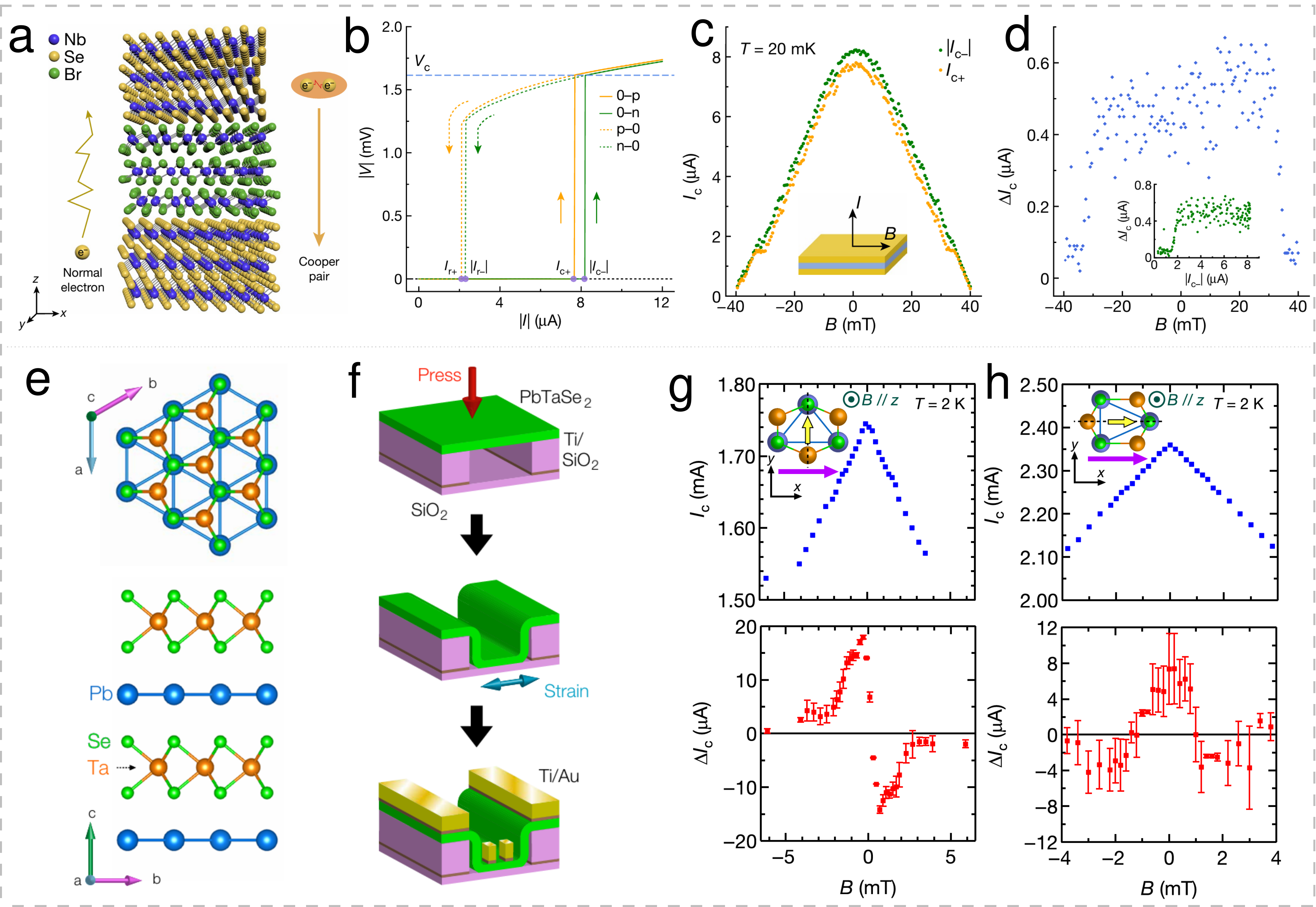}
    \caption{\textbf{Zero-field SDEs in polar structures.}  a) Lattice structure of NbSe$_2$/Nb$_3$Br$_8$/NbSe$_2$ heterostructures, where Nb$_3$Br$_8$ serves as the polar barrier in this JJ. b) Asymmetric $I_\mathrm{c}$ in Nb$_3$Br$_8$ JJs measured at 20~mK. c,d) Field dependence of asymmetric $I_\mathrm{c}$ and $\Delta I_\mathrm{c}$ in Nb$_3$Br$_8$ JJs, respectively, displaying $B$-even features. e) Crystal structure of PbTaSe$_2$. f) Schematic of strain application to PbTaSe$_2$. The polar axis is induced by applying strain along the armchair direction (mirror plane). g) Field dependence of averaged $I_\mathrm{c}$ and $\Delta I_\mathrm{c}$ with strain and current applied along the zigzag direction. In this case, the SDE polarity is $B$-odd. h) Field dependence of averaged $I_\mathrm{c}$ and $\Delta I_\mathrm{c}$ with strain and current applied along the armchair direction. In this case, the SDE polarity is $B$-even. In the insets of (g,h), the yellow arrow denotes the mirror plane and the purple arrow indicates the direction of current and strain. $B$ is applied out-of-plane.   
    (a-d) are reproduced with permission~\cite{wu2022field}, Copyright 2022, The Authors, under exclusive licence to Springer Nature Limited. (e-h) are reproduced with permission~\cite{liu2024superconducting}, Copyright 2024, The Authors, some rights reserved.}
    \label{6}
\end{figure*}

In 2022, Wu $et~al.$ observed zero-field SDEs in a NbSe$_2$/Nb$_3$Br$_8$/NbSe$_2$ vdW heterostructure, as seen in Figure \ref{6}a,b~\cite{wu2022field}. In this device, the odd-layer Nb$_3$Br$_8$ exhibits inversion asymmetry, acting as a polar barrier for the JJs. The zero-field JDEs are thought to be related to the asymmetric tunneling of supercurrent across this barrier. Surprisingly, $I_\mathrm{c}$ and $\Delta I_\mathrm{c}$ demonstrate an unusual symmetric evolution with respect to external $B$ in Figure \ref{6}c,d, distinct from that in the aforementioned diode systems. In this system, the internal TRS-breaking is not obvious and the underlying mechanism for all these observations remains a puzzle. Some researchers proposed that TRS-breaking might be related to the possible existence of strongly correlated flat band of the metallic layer~\cite{zhang2022general}.

Zero-field SDEs are also observed in strained PbTaSe$_2$ films~\cite{liu2024superconducting}. Similar to MoS$_2$, it possesses a trigonal symmetry which is noncentrosymmetric yet nonpolar, as illustrated in Figure \ref{6}e. The structure features mirror planes along the armchair direction. Figure \ref{6}f shows a sketch depicting the strain application. 
By applying tensile strain along the zigzag direction, finite-field SDEs were observed, displaying usual odd behavior with respect to external $B_\mathrm{out}$, as depicted in Figure \ref{6}g. While strain is applied along the armchair direction, zero-field SDEs were detected for current flow in the same direction, exhibiting abnormal field-symmetric behavior, as shown in Figure \ref{6}h. This directional dependence of the SDEs aligns with the crystal symmetry. The $B$-even zero-field SDEs were considered to be associated with electric polarization induced by strain along the mirror plane. Although the exact mechanism remains unclear, the authors argued that TRS-breaking is not a prerequisite for $B$-even zero-field SDEs, rather, the polar structure is essential.

\section{Perspective and outlook}

\noindent
From the fundamental point of view, SDEs serve as powerful tools to probe unconventional phases of matter within the superconducting state. Functionally, SDEs represent groundbreaking approaches to the development of ultra-low energy electronic devices. In contrast with traditional semiconductors, SDEs enable rectification without resistive losses, which offers significant advantages for applications in superconducting logic circuits, switching devices, and quantum bit manipulators. For practical implementations, SDE devices must operate at zero magnetic fields and elevated temperatures while maintaining large rectification and tunability. Although research in this area is still in its infancy, ongoing efforts along these lines are being made to advance the understanding and application of SDEs.

\subsection{SDE rectification efficiency}

\begin{figure*}[h]
    \centering
    \includegraphics[width=1\linewidth]{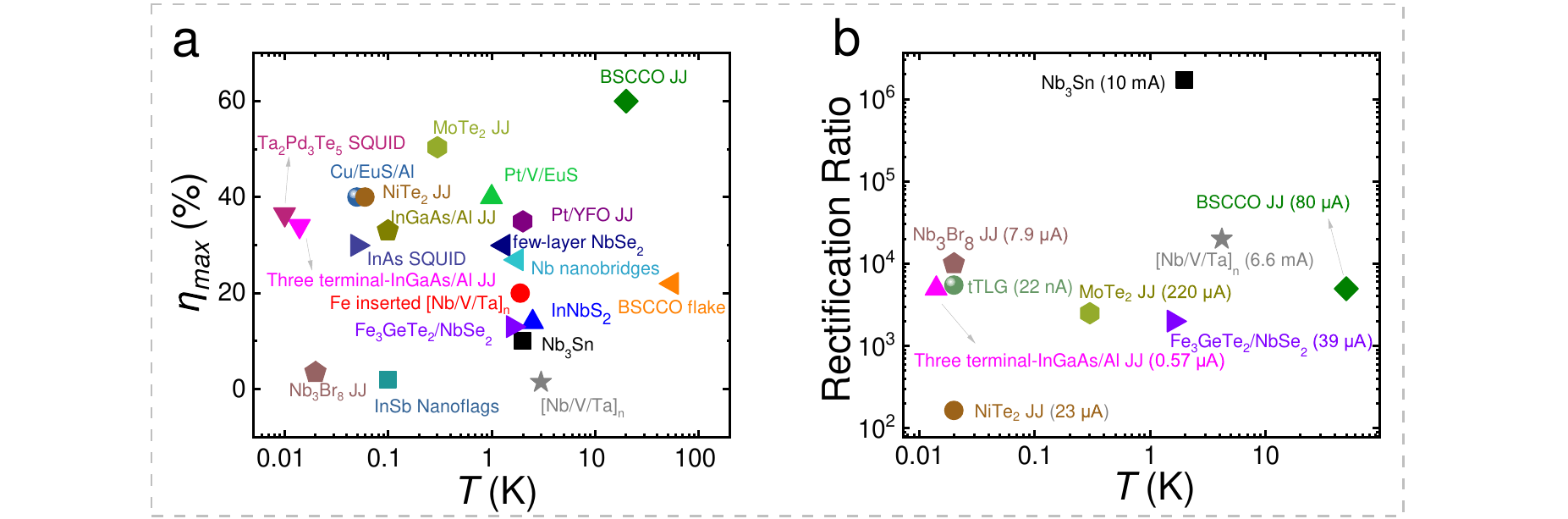}
    \caption{\textbf{SDE performance in various systems.} a) Maximum rectification efficiency across various systems operating at different  temperatures~\cite{chahid2023high,narita2023magnetization,zheng2024larger,gupta2023gate,ghosh2024high,bauriedl2022supercurrent,xiong2024electrical,jeon2022zero,ando2020observation,wu2022field,baumgartner2022supercurrent,strambini2022superconducting,turini2022josephson,pal2022josephson,hou2023ubiquitous,li2024interfering,margineda2023sign,ciaccia2023gate,chen2024edelstein,qi2025high}. b) Corresponding rectification ratio ($V_\mathrm{n}(I_\mathrm{+})/V_\mathrm{s}(I_\mathrm{-})$) for various systems.~\cite{chahid2023high,gupta2023gate,ghosh2024high,xiong2024electrical,ando2020observation,wu2022field,pal2022josephson,chen2024edelstein,lin2022zero}.
}
    \label{7}
\end{figure*}

Rectification efficiency and rectification ratio are key metrics for assessing the performance of SDE devices in practical applications. Rectification efficiency ($\eta$) quantifies the capability of a superconducting diode to facilitate unidirectional flow of supercurrent. In an ideal case, this efficiency achieves $100\%$ if the supercurrent flows preferentially in one direction, while being completely blocked in the opposite direction. Figure \ref{7}a summarizes the maximum rectification efficiency in various devices, including [Nb/V/Ta]$_\mathrm{n}$ \cite{ando2020observation}, NbSe$_2$ thin films\cite{bauriedl2022supercurrent}, and InSb nanosheets\cite{turini2022josephson}. 
Rectification ratio ($V_\mathrm{n}(I_\mathrm{+})/V_\mathrm{s}(I_\mathrm{-})$), as illustrated in  Figure \ref{7}b, is another important metric that characterizes rectification performance. It is defined as the ratio between the voltage in normal state and the noise-level voltage observed in the superconducting state when current flows in the opposite direction. The performance of SDE devices is influenced not only by the intrinsic properties of the materials but also by external factors such as temperature, magnetic field, and geometric engineering.

\subsection{SDEs with large tunability}
Geometry engineering is a powerful approach to manipulate the polarity and rectification efficiency of SDEs. As mentioned previously, techniques such as introducing patterns of micro-nano holes or nanomagnet arrays, can effectively enhance SDE performance~\cite{lyu2021superconducting,li2024unconventional}. Additionally, electrostatic control allows for $in$-$situ$ manipulation of SDE properites, which has been documented in several reports~\cite{gupta2023gate,xiong2024electrical,lin2022zero}. For instance, Figure~\ref{4}d presents non-volatile electrical manipulation of zero-field SDE polarity owning to current-induced spin accumulation, facilitating the development of quantum neuronal transistors~\cite{xiong2024electrical}. Figure \ref{5}c shows the results of tTLG, where both the polarity and the efficiency of the SDEs are adjusted by altering the Fermi level through gating voltages~\cite{lin2022zero}.
Figure~\ref{8}a depicts the gated asymmetric superconducting quantum interference devices (SQUIDs) utilizing a Al/InGaAs/InAs/InGaAs heterojunction, in which the SDE characteristics are directly correlated to the phase shift of the SQUIDs tuned by the asymmetric gate\cite{reinhardt2024link} in Figure~\ref{8}b. Furthermore, the electrical switching of electrical polarization in a SC/ferroelectric/SC heterostructure or even in ferroelectric superconductors~\cite{Rischau2017NP,Jindal2023Nature} could also modulate SDEs, enabling ultrafast non-volatile switching that has yet to be reported. Overall, the expected advances in geometric and electrostatic engineering hold great potential for optimizing SDE functionality.

\subsection{High temperature SDEs}

\begin{figure*}[h]
    \centering
    \includegraphics[width=1\linewidth]{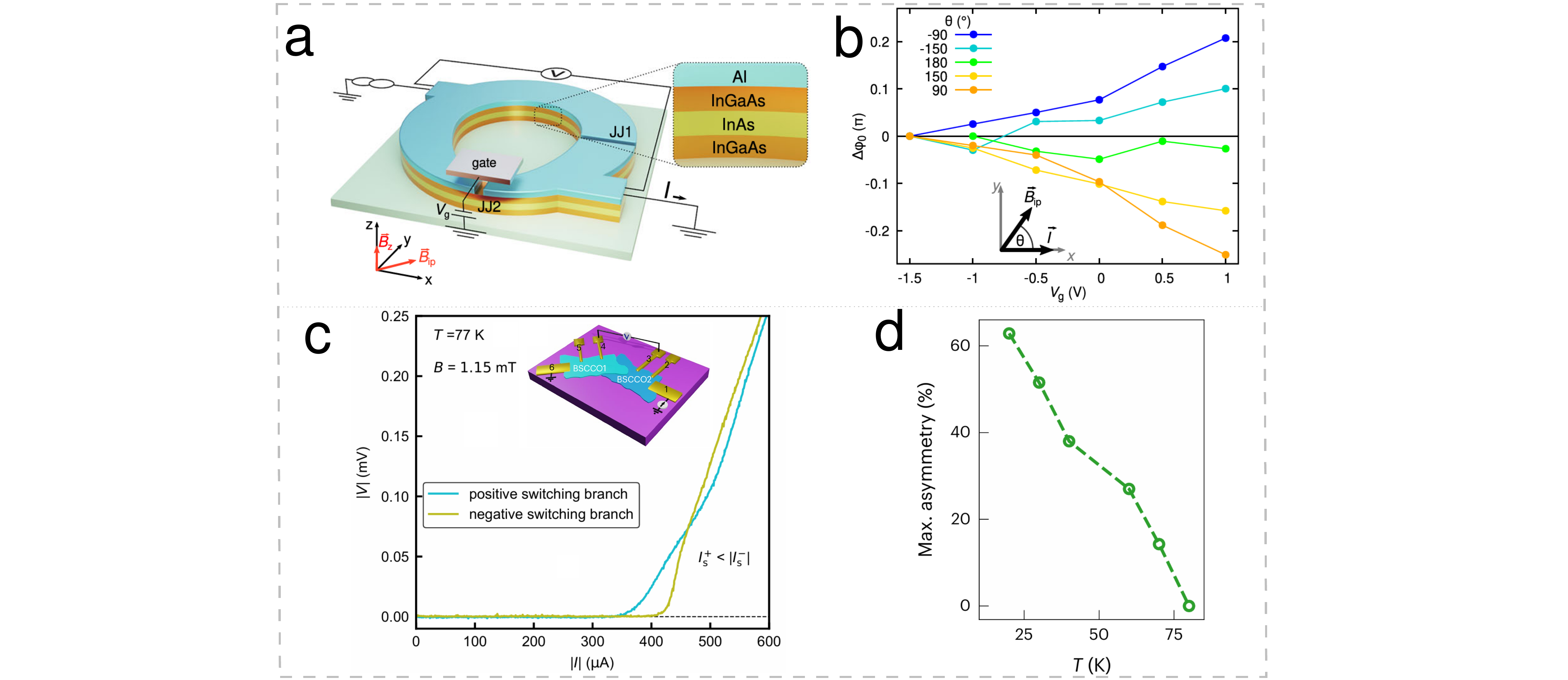}
    \caption{\textbf{SDEs for potential future applications.} a) Schematic illustration of gated asymmetric SQUIDs formed by Al/InGaAs/InAs/InGaAs heterojunction. b) Phase modulation by electrostatic gate at different angles between the direction of external in-plane field and current. c) SDEs observed in high-$T_\mathrm{c}$ cuprate BSCCO-based JJs measured at $T=77$~K and $B=1.15$~mT. The inset shows a sketch of the device. d) Temperature dependence of maximum diode efficiency in BSCCO JJs. (a,b) are reproduced with permission~\cite{reinhardt2024link}, Copyright 2024, The Authors. (c,d) are reproduced with permission~\cite{ghosh2024high}, Copyright 2024, The Authors, under exclusive licence to Springer Nature Limited.
    }
   \label{8}
\end{figure*}

SDE devices, that operate at higher temperatures, are of paramount importance for practical applications. Previous studies on SDEs primarily concentrated on the low-$T_\mathrm{c}$ superconductors, where the effect manifests below the boiling point of liquid helium. Deshmukh $et~al.$ achieved a significant breakthrough by observing the SDEs at 77~K in JJs fabricated from high-$T_\mathrm{c}$ cuprates Bi$_2$Sr$_2$CaCu$_2$O$_{8+x}$ (BSCCO), as shown in Figure \ref{8}c. As the temperature lowers, their performance improves, reaching an efficiency of 60$\%$ at 20~K, as illustrated in Figure \ref{8}d~\cite{ghosh2024high}. This discovery paves the way for the realization of superconducting Josephson circuits operating at liquid nitrogen temperatures, offering a practical pathway to more accessible high temperature superconducting technologies. On the other hand, current-tunable zero-field SDEs observed in approximately $45^{o}$ twisted BSCCO have been used to imply the intrinsic TRS-breaking in the system~\cite{zhao2023time}. Intriguingly, $B$-even SDEs have been observed in BSCCO flake devices, persisting up to 75~K with an efficiency of 22$\%$ at 53~K, and the authors proposed that this may be related to loop currents causing the breaking TRS and IRS~\cite{qi2025high}.

~\\
In summary, practical applications of SDE technology remain a distant goal, primarily constrained by several critical challenges. First of all, the availability of superconductors capable of zero-field SDEs is still limited, combined with the challenge of integrating them with the existing electronics. Secondly, high-frequency rectifiers essential for practical applications encounter obstacles such as thermal management and noise control~\cite{chahid2023high}. Finally, there remains a gap between theoretical models and experimental validation. Nonetheless, the ongoing progress in materials science, experimental techniques, and theoretical frameworks will steadily push SDEs closer to transformative applications for next-generation superconducting circuit technology, particularly in low-power electronics and quantum computing.

\clearpage

% Secondly, the fabrication process of SDE devices is complex, particularly when integrating them with existing electronic systems, where issues of compatibility and design intricacies arise. Additionally, high frequency rectifiers encounter challenges in thermal management and noise control. Lastly, there remains a gap between theoretical models and experimental validation.} However, we expect that the ongoing progress in materials science, experimental techniques, and theoretical frameworks will steadily push SDEs closer to transformative applications for next-generation superconducting circuit technology, particularly in low-power electronics and quantum computing.

\noindent
\\
\noindent
\textbf{Acknowledgements} \par 
\noindent
This research is supported by ``Pioneer'' and ``Leading Goose'' R$\&$D Program of Zhejiang under Grant 2024SDXHDX0007 and
Zhejiang Provincial Natural Science Foundation of China for Distinguished Young Scholars under Grant No. LR23A040001. 
This research is supported by National Natural Science Foundation of China under Grant No. 12474131. 
This research is supported by the Research Center for Industries of the Future (RCIF) at Westlake University under Award No. WU2023C009.

~\\
\noindent
\textbf{Author Contribution} \par 
\noindent
{\dag}\verb|Equal contributions.|\\
J. M. and R. Z. contributed equally to this review.

\medskip

\clearpage

\bibliographystyle{MSP-1}
\bibliography{SDE}

\end{document}